\newcommand{\D}{\mathrm{d}}
\newcommand{\E}{\mathrm{e}}
\newcommand{\I}{\mathrm{i}}
\newcommand{\re}{\mathrm{Re}}

\newcommand{\bscco}{Bi$_2$Sr$_2$CaCu$_2$O$_{8+ \delta}$}

\newcommand{\lsco}{La$_{2-x}$Sr$_x$CuO$_4$}

\newcommand{\hgbco}{HgBa$_2$CuO$_{4+\delta}$}
\newcommand{\ybco}{YBa$_2$Cu$_3$O$_{6 + y}$}

\newcommand{\ybcou}{YBa$_2$Cu$_3$O$_{6.333}$}
\newcommand{\ybcoOI}{YBa$_2$Cu$_3$O$_{6.99}$}
\newcommand{\ybcoOII}{YBa$_2$Cu$_3$O$_{6.50}$}

\newcommand{\tc}{$T_c$}
\newcommand{\td}{$T_d$}

\newcommand{\dwave}{$d_{x^2 - y^2}$}

\documentclass[aps, prb, twocolumn, floatfix, amssymb, superscriptaddress]{revtex4}

\usepackage{graphicx}% Include figure files
\usepackage{dcolumn}% Align table columns on decimal point
\usepackage{bm}% bold math
\usepackage{amssymb,amsmath}
\usepackage{color}
\usepackage{array}
\usepackage{multirow}
\usepackage{booktabs}
\begin{document}

%\preprint{APS/123-QED}

\title{Stability of nodal quasiparticles in underdoped \ybco\ \\probed by penetration depth and microwave spectroscopy}% Force line breaks with \\
\author{W.~A.~Huttema}
\affiliation{Department of Physics, Simon Fraser University, Burnaby, BC, V5A 1S6, Canada}
\author{J.~S.~Bobowski}
\affiliation{Department of Physics and Astronomy, University of British Columbia, Vancouver, BC, V6T 1Z1, Canada}
\author{P.~J.~Turner}
\affiliation{Department of Physics, Simon Fraser University, Burnaby, BC, V5A 1S6, Canada}
\author{Ruixing Liang}
\author{W.~N.~Hardy}
\author{D.~A.~Bonn}
\affiliation{Department of Physics and Astronomy, University of British Columbia, Vancouver, BC, V6T 1Z1, Canada}
\author{D.~M.~Broun}
\affiliation{Department of Physics, Simon Fraser University, Burnaby, BC, V5A 1S6, Canada}

\date{\today}% It is always \today, today,

\begin{abstract}
High resolution measurements of superfluid density $\rho_s(T)$ and broadband quasiparticle conductivity $\sigma_1(\Omega)$ have been used to probe the low energy excitation spectrum of nodal quasiparticles in underdoped \ybco.  Penetration depth $\lambda(T)$ is measured to temperatures as low as 0.05~K. $\sigma_1(\Omega)$ is measured from 0.1 to 20~GHz and is a direct probe of zero-energy quasiparticles. The data are compared with predictions for a number of theoretical scenarios that compete with or otherwise modify pure  $d_{x^2 - y^2}$ superconductivity, in particular commensurate and incommensurate spin and charge density waves; $d_{x^2 - y^2} +\I s$ and $d_{x^2 - y^2} + \I d_{xy}$ superconductivity; circulating current phases; and the BCS--BEC crossover.  We conclude that the data are consistent with a pure $d_{x^2 - y^2}$ state in the presence of a small amount of strong scattering disorder, and are able to rule out most candidate competing states either completely, or to a level set by the energy scale of the disorder, \td~$ \sim 4 $~K.  Commensurate spin and charge density orders, however, are not expected to alter the nodal spectrum and therefore cannot be excluded.
\end{abstract}

\maketitle

\section{Introduction}

%Outstanding: ARPES TRSB;

The physics of the cuprate high temperature superconductors is that of strong Coulomb repulsion in nearly half-filled CuO$_2$ planes.\cite{orenstein00,bonn06}  As charge carriers are doped into these materials, the two most prominent electronic states are the antiferromagnetic (AFM) Mott insulator and the $d$-wave superconductor.   While the AFM and the optimal-to-overdoped superconductor appear to be well understood, the physics of the underdoped part of the phase diagram that lies between them remains firmly incompatible  with standard theory. The most prominent feature of this region is a pseudogap that suppresses low energy spin and charge fluctuations and persists above the superconducting transition to a temperature $T^\ast$.\cite{warren89, orenstein90, homes93}  The pseudogap temperature is highest close to the Mott insulator and decreases monotonically as doping, $p$, is increased towards optimal doping. Identifying the nature of the pseudogap state remains a difficult and open problem.

States of matter are characterized by their symmetries and their low energy excitation spectra.  $d$-wave superconductivity, for instance, breaks four-fold rotational symmetry and is distinguished by the presence of nodal quasiparticles with a characteristic linear energy spectrum.  The $d$-wave state in the cuprates was first identified from observations of a linear temperature dependence of penetration depth, $\lambda$, and superfluid density, $\rho_s \equiv 1/\lambda^2$.\cite{hardy93} The ability of superfluid density to couple directly to itinerant electronic degrees of freedom gives it the potential to be a sensitive thermodynamic probe of pseudogap physics, with many candidate states expected to leave characteristic signatures in the low energy quasiparticle spectrum.  Here we search for these signatures using high resolution measurements of penetration depth and broadband quasiparticle conductivity, made on very clean crystals of underdoped \ybco.

There have been a wide range of proposals put forward to explain the cuprate pseudogap.  In one important category, strong pair correlations are already built into the normal state.  This scenario has its roots in Anderson's resonating-valence-bond spin liquid,\cite{anderson87} and the idea that pair correlations emerge directly from the Mott insulator remains a compelling proposition. The `gossamer superconductor' --- a BCS wavefunction in which double occupancy has been heavily suppressed --- typifies this approach and may provide a useful representation of the underdoped electronic state.\cite{laughlin06}  The implication for the phase diagram is that $T^\ast$ marks the formation of tightly bound Cooper pairs, with low phase stiffness and strong quantum and thermal phase fluctuations heavily suppressing $T_c$.\cite{emery95,franz01,herbut02,franz02,herbut02a,herbut05}  At temperatures not too far above the superconducting transition, the idea of pre-existing pairs finds support from a number of experiments: terahertz spectroscopy reveals a finite phase-stiffness;\cite{corson99} Nernst-effect measurements appear to detect the phase-slip voltage from thermally diffusing vortices;\cite{xu00,wang02} high-field magnetometry reveals excess diamagnetism;\cite{wang05} and STM\cite{gomes07} and $\mu$SR\cite{sonier08} detect what appear to be droplets of precursor superconductor.  Related to this, the theory of the BCS to Bose--Einstein condensate (BEC) crossover makes a prediction that can be tested here: a  $T^{3/2}$ power law in $\rho_s(T)$, due to the direct thermal excitation of bound Cooper pairs.\cite{chen98,chen06}

Another class of proposals seeks to explain the pseudogap in terms of competing orders and quantum criticality.  In such a scenario,  $T^\ast(p)$ marks the boundary of a distinct thermodynamic phase; must be accompanied by a broken symmetry; and goes to zero at a quantum critical point within the superconducting phase.  This idea was initially motivated by the observation near optimal doping of so-called marginal Fermi liquid behaviour,\cite{varma89} in which unusual power laws in resistivity $\rho(T)$, optical conductivity $\sigma_1(\Omega)$ and other physical quantities could be understood in terms of scattering from a scale-invariant fluctuation spectrum, as would be expected near a zero-temperature critical point.\cite{sachdev00}  On crossing $T^\ast$, these fluctuations should generically condense to form the broken symmetry state of the pseudogap phase. While there is evidence of an AFM quantum critical point in electron-doped materials,\cite{dagan04} the situation on the hole-doped side is much less clear. Identification of a particular competing order that appears at  $T^\ast(p)$ would have strong implications not just for the pseudogap, but for the origin of non-Fermi-liquid behaviour elsewhere in the cuprate phase diagram. 

Competing orders are in fact prevalent in the cuprates, in part as a result of the extreme sensitivity of the doped Mott insulator to perturbations.\cite{sachdev03} Outside the AFM phase, long-range magnetic order is replaced by glassy spin correlations,\cite{kiefl89,weidinger89,panagopoulos02} although this short-range magnetism is likely a response to chemical disorder.\cite{sachdev03,kivelson03} Neutron scattering experiments on \lsco\ have revealed incommensurate spin correlations in superconducting samples\cite{thurston89, cheong91,mason92} that were later identified as stripe ordering of spins and holes.\cite{tranquada95,tranquada97} Stripe correlations appear to be widespread in the underdoped cuprates, and are particularly strong near $p = \frac{1}{8}$ doping.\cite{kivelson03}  In applied field, the suppression of superconductivity in vortex cores\cite{zhang97,arovas97} leads to co-existing superconductivity and spin-density-wave order.\cite{lake01,lake02,demler01,kivelson02,sonier07}  Scanning tunneling spectroscopy of the vortex cores reveals that this is accompanied by prominent checkerboard charge-density order.\cite{hoffman02}  Similar four-lattice-constant modulations of the density of states are seen in zero field at various points in the phase diagram.\cite{howald03,hanaguri04} For the most part, these competing orders occur in narrow ranges of doping; or in particular materials; or in response to external perturbations such as point disorder or applied magnetic field.  While they attest to the complexity of the doped Mott insulator,\cite{sachdev03} they offer only  hints at the physics underlying the formation of the pseudogap.  

The lack of compatibility of the observed ordered states with $T^\ast(p)$ has led to interest in `hidden orders', in which the broken symmetry is subtle and difficult to detect with standard scattering experiments.  Proposals include circulating current phases that preserve translational symmetry\cite{varma97,varma99,varma06} and orbital antiferromagnetism, for example the $d$-density wave state (DDW).\cite{chakravarty01}  Interestingly, a set of recent experiments now appears to have detected signatures of one or more of these phases.  $\mu$SR\cite{sonier01} and polar Kerr effect\cite{xia08} have established the onset of time-reversal-symmetry breaking (TRSB) at $T^\ast(p)$ in \ybco, but the signals are extremely weak.  It has been suggested that a variant of the DDW, the $d_{xy} + \I d_{x^2 - y^2}$ density wave,\cite{tewari08} would contain a subdominant but macroscopic TRSB component and be consistent with the small magnitude of the observed effects.   Spin-polarized neutron scattering on \ybco\ has detected weak signatures of a novel magnetic order that preserves translational symmetry,\cite{fauque06} and has a form consistent with the $\Theta_{II}$ circulating current phase proposed by Varma,\cite{varma06} shown in schematic form in the inset of Fig~\ref{thetaIIcurrents}.  The detailed picture is complicated by the presence of  an in-plane component of magnetic moment, although it has been suggested that this could arise from orbital currents that circulate through apical oxygens while preserving the $\Theta_{II}$ symmetry.\cite{weber08}  It also remains to be seen how ubiquitous the effects are: $\mu$SR experiments on \lsco\ have so far failed to observe TRSB,\cite{macdougall08} but have not yet been carried out with the same sensitivity as Ref.~\onlinecite{sonier01}.  In contrast, new neutron scattering experiments\cite{li08} on \hgbco\ have detected the same type of $\Theta_{II}$ magnetic order seen in \ybco.  As we will discuss in more detail below, this type of order has a strong effect on the low energy states of the superconductor, and should be highly visible in measurements of $\rho_s(T)$.

Finally, there have been suggestions that pure $d_{x^2 - y^2}$ superconductivity may compete with superconducting states of different symmetry,\cite{laughlin98, balatsky00,vojta00} motivated in part by reports of anomalously large inelastic scattering of nodal quasiparticles \emph{below} $T_c$.\cite{valla99,corson99}  This critical-like scattering has been shown to be compatible with a quantum phase transition to a $d_{x^2 - y^2} + \I s$ or $d_{x^2 - y^2} + \I d_{xy}$ state.\cite{vojta00} To date, there is a limited amount of direct experimental evidence in support of such phases\cite{krishana97,dagan01,daghero03} --- here we use measurements of superfluid density to place tight constraints on the existence of such states in \ybco.

This paper is organized as follows.  In Sec.~\ref{penetrationdepththeory} we show how measurements of penetration depth and broadband microwave conductivity can together be used as a probe of the quasiparticle excitation spectrum and the structure of the superconducting energy gap.  In Sec.~\ref{competingorders} we catalog how different competing orders affect the superfluid density, including the effect of disorder.  In Sec.~\ref{experiment} we introduce the experimental methods used to measure superfluid density and broadband microwave conductivity.  Results are presented and discussed in Sec.~\ref{results}, followed by a summary of our conclusions in Sec.~\ref{conclusions}.  Appendix~\ref{appendix} presents analytic results for the effect of disorder on $d_{x^2 - y^2}$, $d_{x^2 - y^2} + \I d_{xy}$ and $d_{x^2 - y^2} + \I s$-type superconductors with isotropic Fermi surfaces, and shows how this eventually blurs the distinction between $d_{x^2 - y^2}$ and $d_{x^2 - y^2} + \I d_{xy}$ states.

\section{Penetration Depth and Microwave Conductivity}\label{penetrationdepththeory}

Microwave experiments can be used to probe the low-energy excitation spectrum of a superconductor in two ways: through the temperature dependence of the penetration depth $\lambda(T)$; and from broadband measurements of the oscillator strength in the finite-frequency quasiparticle conductivity spectrum $\sigma_1(\Omega, T)$.  The theory of penetration depth and microwave conductivity of unconventional superconductors has been developed in great detail,\cite{nam67,pethick86,hirschfeld88,prohammer91,schachinger03,hirschfeld93,hirschfeld93a,borkowski94,hirschfeld94} but useful insights about low-lying excitations can be obtained from the weak-coupling BCS theory.  For the case of an isotropic Fermi surface, which should be adequate for describing the low-lying excitations in the cuprates, $\lambda(T)$ is given by\cite{tinkham,waldram}  
\begin{align}
\frac{\lambda^2_0}{\lambda^2(T)}   & =   1 -  \int_{-\infty}^\infty  \!\!\!\!\D \omega\left(\!\!-\frac{\partial f}{\partial \omega}\!\right) N(\omega)\label{lambdaone}\\ 
 & =  \tfrac{1}{2} \int_{-\infty}^\infty  \!\!\!\!\D \omega\, \tanh\left(\frac{\omega}{2 k_B T}\right)\frac{\partial N(\omega)}{\partial \omega}\;.\label{lambdatwo}
\end{align}
Here $\lambda_0$ is the zero-temperature penetration depth in the \emph{absence} of disorder and competing phases, and $f(\omega/T)$ is the Fermi function.  A Sommerfeld expansion reveals the direct connection between $\lambda(T)$ and the normalized density of quasiparticle states $N(\omega)$: if $N(\omega) = N_0 + N_1 \omega + \frac{1}{2}N_2 \omega^2 + ...$ then \mbox{$\lambda^2_0/\lambda^2(T) = 1 - N_0 - 2 \ln 2 N_1 k_\mathrm{B} T - \frac{\pi^2}{6}N_2 ( k_\mathrm{B} T)^2 - ...\;$}  The residual density of states (DOS) $N_0$ represents zero-energy excitations, which arise in a superconductor either from impurity pair-breaking, or from certain types of competing order, notably the $\Theta_{II}$-type circulating current phase\cite{varma06, berg08}. Note that $N_0$ does not appear in the temperature dependence of $\lambda$, but instead results in a deviation of $\lambda(T\!\!\to \!\!0)$ from $\lambda_0$.  This shift in penetration depth is difficult to resolve experimentally, because $\lambda_0$ is neither known \emph{a priori}, nor can the absolute value of $\lambda(T \to 0)$ usually be measured with sufficient accuracy.  However, a direct determination of $N_0$ can be obtained from the uncondensed weight spectral in the quasiparticle conductivity $\sigma_1(\Omega, T)$. From the oscillator strength sum rule,
\begin{equation}
N_0 = \tfrac{2}{\pi} \mu_0 \lambda_0^2 \int_0^{\Omega_c} \!\!\!\!\sigma_1(\Omega, T\!\to \!0)\, \D \Omega\; ,
\end{equation}
where $\Omega_c$ is a frequency cut-off chosen to capture the oscillator strength of the conduction electrons only.

In a clean-limit BCS superconductor, $N(\omega)$ is determined by the k-space structure of the superconducting order parameter $\Delta_\mathbf{k}$: $N(\omega) = \re \big\langle \omega /\sqrt{\omega^2 - \Delta_\mathbf{k}^2}\big\rangle_\mathrm{FS}$, where $\big\langle ... \big\rangle_\mathrm{FS}$ denotes a Fermi surface average. This makes $\rho_s(T)$ a sensitive probe of order parameter symmery.  In particular, for a $d$-wave superconductor in two dimensions, the linear dispersion of $\Delta_\mathbf{k}$ about the gap nodes leads to $N(\omega) \propto \omega$ and $\Delta \rho_s(T) \propto T$.  An $s$-wave superconductor, by constrast, usually has a finite energy gap and shows activated behaviour, $\rho_s(T) \propto \exp(-\Delta_\mathrm{min}/k_B T)$, where $\Delta_\mathrm{min}$ is the minimum of the energy gap on the Fermi surface. The effect of impurity scattering on $N(\omega)$ and $\rho_s(T)$ is important and is reviewed in Appendix~\ref{appendix}, where we give analytic results for $d_{x^2 - y^2}$, $d_{x^2 - y^2} + \I d_{xy}$ and $d_{x^2 - y^2} + \I s$ superconductors with isotropic Fermi surfaces in the presence of point defects. The main effect of disorder is for the quasiparticles to acquire a lifetime, the magnitude and energy dependence of which depend on the concentration and scattering strength of the defects.  Near the unitarity limit, scattering leads to a zero-energy resonance that overlaps with the continuum of quasiparticle states in the $d_{x^2 - y^2}$-wave superconductor, resulting in a residual density of states in $N(\omega)$ and a crossover to $T^2$ behaviour in $\rho_s(T)$.  This also happens for the $d_{x^2 - y^2} + \I d_{xy}$ superconductor, despite there initially being a finite gap in the excitation spectrum.  As a result, above a certain level of disorder, $d_{x^2 - y^2}$ and $d_{x^2 - y^2} + \I d_{xy}$ states become impossible to tell apart using microwave spectroscopy.  In Fig.~\ref{disordercrossover} we show how the distinction is lost when the energy scale of the disorder, $k_B T_d \gtrsim \Delta_{d_{xy}}$. The  $d_{x^2 - y^2} + \I s$ superconductor is different in this respect: nonmagnetic scatterers do not cause pair breaking at low energies, and the gap in the spectrum is robust.

\section{Superfluid density and competing orders}\label{competingorders}

As a sensitive thermodynamic probe that couples directly to current-carrying excitations, measurements of superfluid density are well suited to detecting changes in the nodal quasiparticle spectrum arising from competing orders and other physics.  A number of authors have investigated these effects theoretically.  Sharapov and Carbotte have performed calculations for a $d_{x^2 - y^2} + \I d_{xy}$ order parameter and for \emph{incommensurate} spin density waves that nest the nodal points (nested SDW), obtaining analytic results for $\rho_s(T \to 0)$ and its leading temperature corrections.\cite{sharapov06}  In the absence of disorder they find that both the nested SDW and the $d_{x^2 - y^2} + \I d_{xy}$ superconductor have a finite gap everywhere on the Fermi surface, leading to activated exponential behaviour $\rho_s(T) \sim \exp(-\Delta'/k_B T)$, where $\Delta'$ is the magnitude of the SDW or $d_{xy}$ gap.  However, nested SDW orders compete for Fermi surface, removing nodal states from the $T = 0$ condensate.  In contrast, a transition to a clean $d_{x^2 - y^2} + \I d_{xy}$ state leaves $\rho_s(T \to 0)$ unchanged.  Unfortunately, this distinction is difficult to detect experimentally, for reasons discussed in Sec.~\ref{penetrationdepththeory}.  In the presence of disorder, both the nested SDW and $d_{x^2 - y^2} + \I d_{xy}$ states develop a leading quadratic temperature dependence, $\rho_s \sim T^2$, similar to that of a dirty $d_{x^2 - y^2}$ superconductor.  However, an experimentally detectable difference now arises: pair-breaking in the $d_{x^2 - y^2} + \I d_{xy}$ state is accompanied by  zero-energy quasiparticles, whereas the disordered SDW continues to remove low energy states \emph{without} creating a residual DOS.  Atkinson has studied the competition between nested, incommensurate SDW and $d_{x^2 - y^2}$ superconductivity numerically and finds broadly similar results,\cite{atkinson07} pointing out that on the basis of the temperature dependence of $\rho_s$ alone, the effect of disordered magnetism cannot be distinguished from dirty but pure $d$-wave superconductivity. He shows that the suppression of zero-temperature superfluid density in the nested SDW case arises because nodal Cooper pairs cease to carry a well-defined current.  Modre \emph{et al.}\ have  studied the $d_{x^2 - y^2} + \I s$ pairing state, which also has a finite energy gap and activated behaviour in $\rho_s(T)$ at low temperature.\cite{modre98}  In contrast to the $d_{x^2 - y^2} + \I d_{xy}$ case, the $d_{x^2 - y^2} + \I s$ gap is stable in the presence of disorder of \emph{any} strength.  In Appendix~\ref{appendix} we show how this arises from impurity renormalization of the $s$-wave gap component.

\begin{figure}[t]
\begin{center}
\includegraphics[width=83mm]{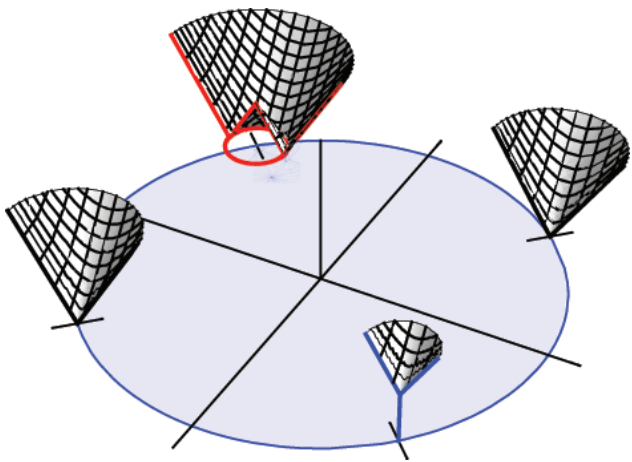}
\caption{(color online).  The presence of a perturbation of the form Eq.~\ref{thetaII}, from the $\Theta_{II}$-type circulating currents shown in Fig.~\ref{thetaIIcurrents}, modifies the nodal spectrum of the $d_{x^2 - y^2}$ superconductor in a characteristic way: one node is shifted up in energy by $\approx 4 \Delta_\mathrm{cc}$, one is shifted down, and two are unperturbed.} 
\label{thetaIIspectrum}
\end{center}
\end{figure}

\begin{figure}[t]
\begin{center}
\includegraphics[width=65mm]{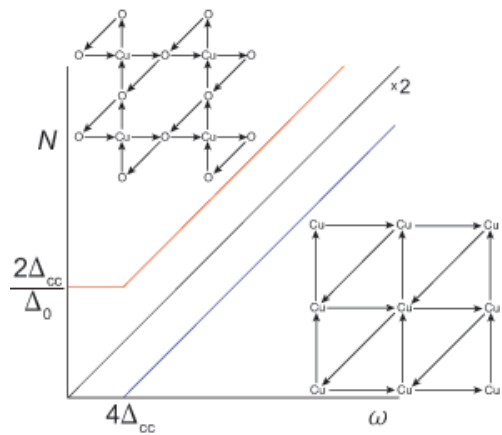}
\caption{Individual nodal contributions to the density of states $N(\omega)$ from a circulating current perturbation of the form Eq.~\ref{thetaII}.  Inset, upper left: the $\Theta_{II}$ circulating current pattern proposed in Ref.~\onlinecite{varma06}.  Inset, lower right: an equivalent current pattern within a one-band model of the CuO$_2$ planes.\cite{berg08}} 
\label{thetaIIcurrents}
\end{center}
\end{figure}

Berg \emph{et al.}\ have studied the stability of the nodal quasiparticle spectrum in the presence of \emph{commensurate} competing orders of all types.\cite{berg08}  For commensurate perturbations that do not nest the nodal points, they prove that if the perturbation is invariant under time reversal or time reversal followed by a lattice translation, the nodal spectrum is stable.  While it remains uncertain whether the converse holds in general, they examine several important cases in which the nodal spectrum breaks down, including certain stripe-like arrangements of spin and charge density, and the $\Theta_{II}$ circulating-current phase that has been detected by neutron scattering in \ybco\ and \hgbco.  Confining themselves to a one-band model of the CuO$_2$ planes, Berg \emph{et al.}\ have used the simpler arrangement of orbital currents shown in Fig.~\ref{thetaIIcurrents}, which is equivalent to the $\Theta_{II}$ state in the more complicated three-band Cu--O lattice of Ref.~\onlinecite{varma06}.  For a perturbation to the pure $d_{x^2 - y^2}$ superconductor of the form 
\begin{equation}
W = - \I \Delta_\mathrm{cc} \left\{\sum_{\mathbf{r}\mathbf{r}'\sigma} \eta_{\mathbf{r}\mathbf{r}'}c^\dagger_{\mathbf{r}\sigma} c_{\mathbf{r}'\sigma} + \mbox{h.c.} \right\}\;,\label{thetaII}
\end{equation}
 where $\eta_{\mathbf{r}\mathbf{r}'} = \pm 1$ is determined by the direction of the bond currents in Fig.~\ref{thetaIIcurrents}, they find excitation energies \mbox{$E_\mathbf{k}^\pm = E_\mathbf{k}^0 + 2 \Delta_\mathrm{cc}\left\{ \sin(k_x a) - \sin(k_y a) + \sin[(k_y - k_x) a] \right\}$}. Here $E_\mathbf{k}^0$ is the unperturbed $d$-wave spectrum and $a$ is the lattice spacing.  The perturbed nodal spectrum for the $\Theta_{II}$ state is plotted in Fig.~\ref{thetaIIspectrum}.  The effect of the circulating currents is similar to the Doppler shift from a uniform current applied along a diagonal direction: one node shifts up in energy by $\approx 4 \Delta_\mathrm{cc}$, one node shifts down, and two are unperturbed.  The individual and combined contributions to the low energy DOS are plotted in Fig.~\ref{thetaIIcurrents}.  The net effect on $N(\omega)$ is a finite residual DOS $\approx 2\Delta_\mathrm{cc}/\Delta_0$, and a kink at $\omega \approx 4 \Delta_\mathrm{cc}$ above which the linear energy dependence doubles in slope. In the clean limit, the superfluid density can be obtained from Eqs.~\ref{lambdaone} and \ref{lambdatwo} and is plotted in Table~\ref{competingordertable}.  The limiting low temperature behaviour of $\rho_s(T)$ is linear, arising from excitations near the two unperturbed nodes.  At a temperature of order $4 \Delta_\mathrm{cc}/k_B$, $\rho_s(T)$ crosses over to a second linear regime in which all four nodes contribute and the temperature slope doubles.  In a clean sample, the combination of a residual DOS and a kink in $\rho_s(T)$ separating two linear regimes should be easily observable in experiments.  Calculations in the presence of disorder have not been carried out, but we expect strong scattering impurities to induce additional residual DOS and to cause a crossover to $T^2$ behaviour in $\rho_s(T)$, as is seen in $d_{x^2 - y^2}$ and $d_{x^2 - y^2} + \I d_{xy}$ superconductors.  Although disorder will mask the effect of circulating currents when the crossover temperature $T_d \gtrsim 4 \Delta_\mathrm{cc}/k_B$, it is expected that tight limits on the size of $\Delta_\mathrm{cc}$ can nevertheless be placed, either using $\rho_s(T)$ or from the magnitude of the uncondensed spectral weight in $\sigma_1(\Omega, T \to 0)$.

The effect of competing orders on $\rho_s(T)$ and the residual DOS is summarized in Table~\ref{competingordertable}.  The SDW results are for the case of ordering wavevectors that nest the nodal points.  The response to nested charge density waves is expected to be broadly similar, with the opening of a finite nodal gap that competes for Fermi surface.

\begin{table*}[t]
\begin{tabular}[t]{|c|c|ccccc|}
\hline

\multicolumn{2}{|c|}{} & \rule[-2mm]{0mm}{6mm}$d_{x^2 - y^2}$ & $\Theta_{II}$ current loops & $d_{x^2 - y^2} + \I s$ & $d_{x^2 - y^2} + \I d_{xy}$  & nested SDW\\  
\hline

\multicolumn{2}{|c|}{\raisebox{10 mm}{\renewcommand{\arraystretch}{1.0}\begin{tabular}[t]{c}
\includegraphics[width=20mm]{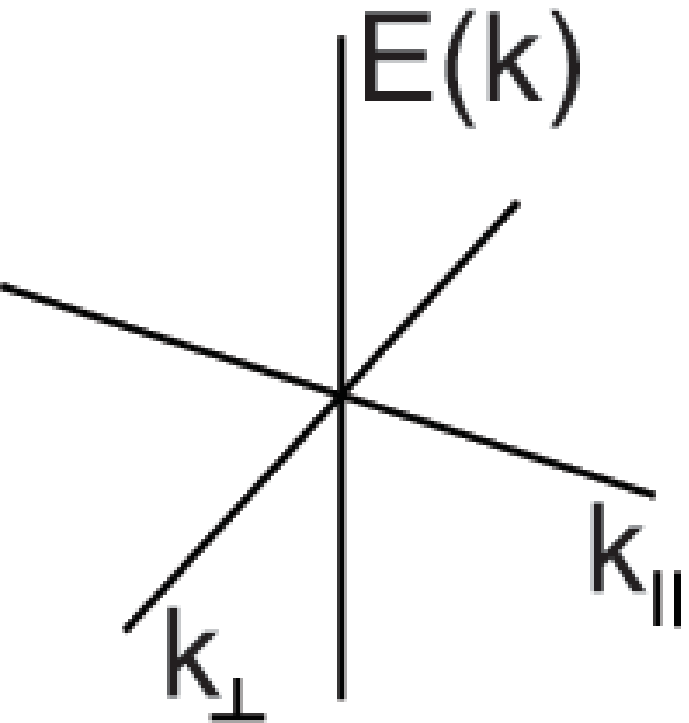}\\
energy\\
spectrum\\
\end{tabular}}}&
 \rule[0mm]{0mm}{39mm} \includegraphics[width=29mm]{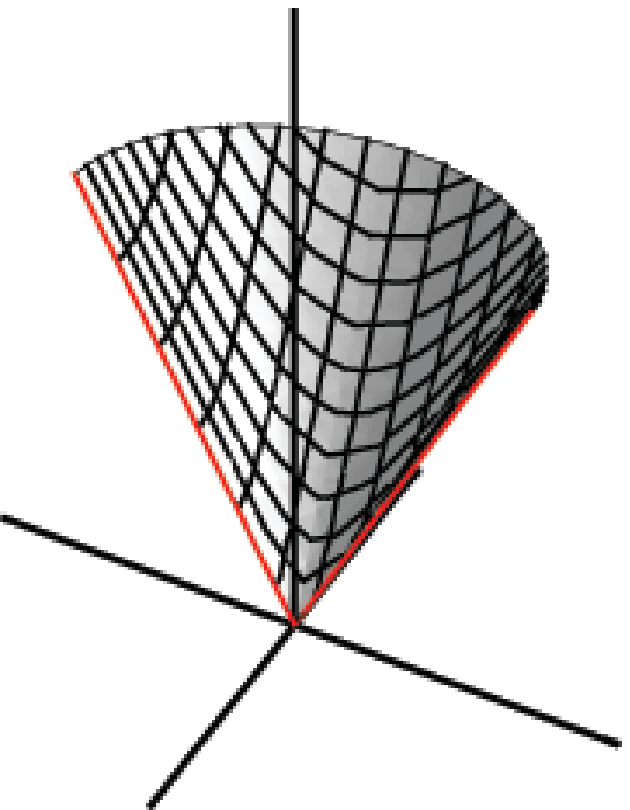} &
\includegraphics[width=29mm]{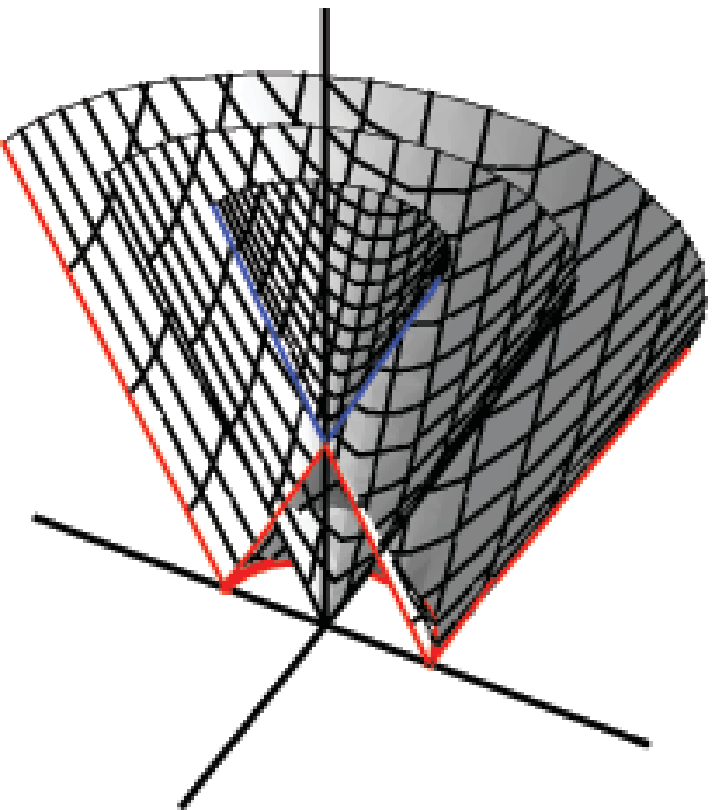} &
\includegraphics[width=29mm]{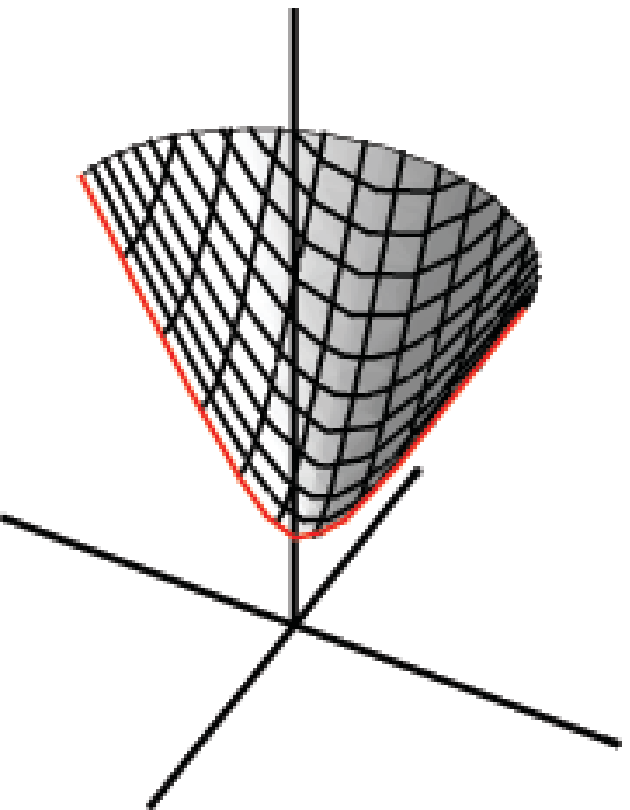}  &
\includegraphics[width=29mm]{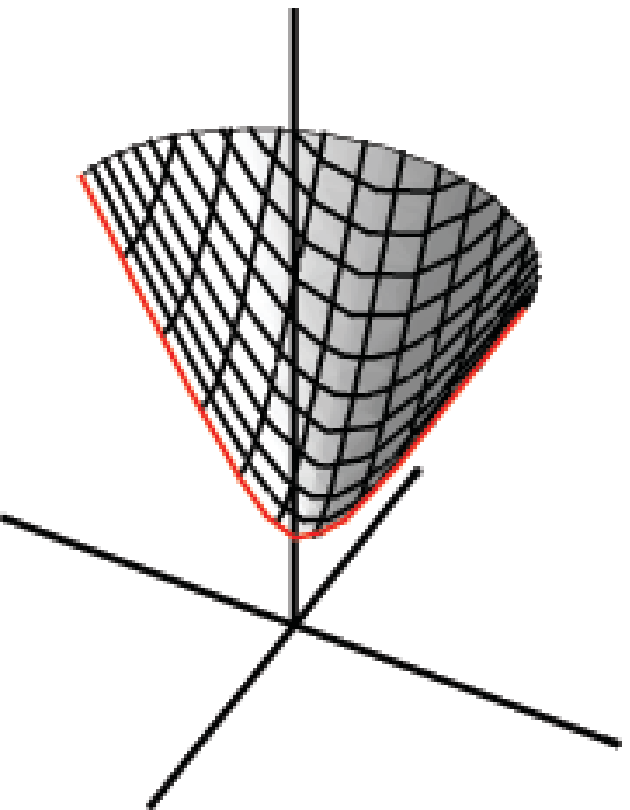} &
\includegraphics[width=29mm]{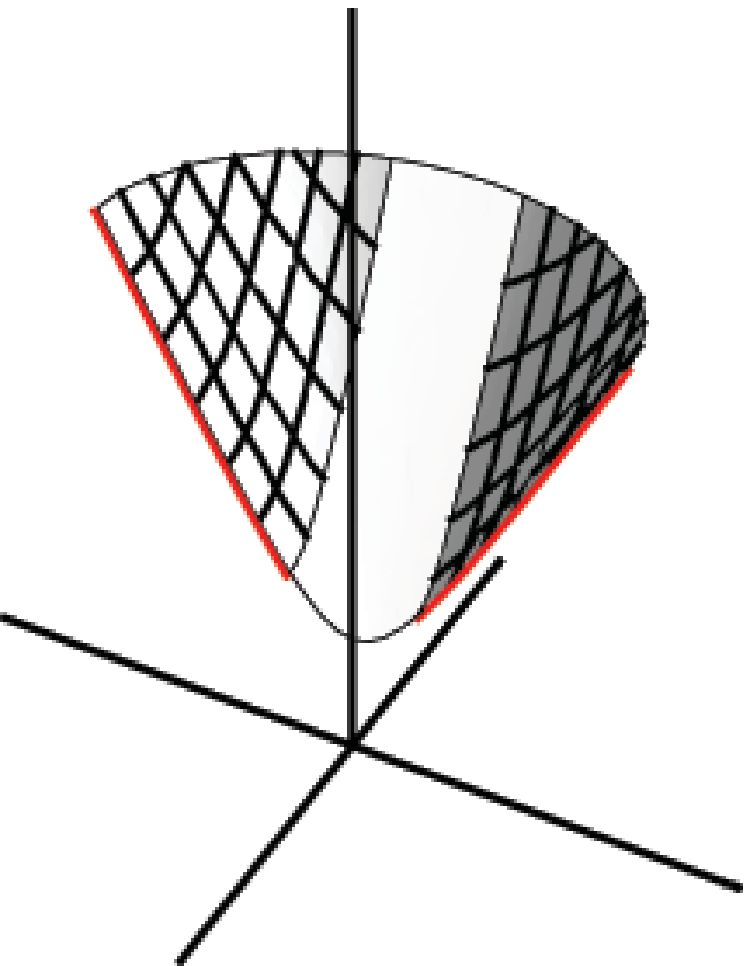} \\ 
\hline

\multicolumn{2}{|c|}{\raisebox{17 mm}{\renewcommand{\arraystretch}{1.0}\begin{tabular}[t]{c}
DOS\\
\textcolor{blue}{clean -- solid}\\
\textcolor{red}{dirty -- dashed}\\
\end{tabular}}}&
 \rule[-2mm]{0mm}{32mm} \includegraphics[width=29mm]{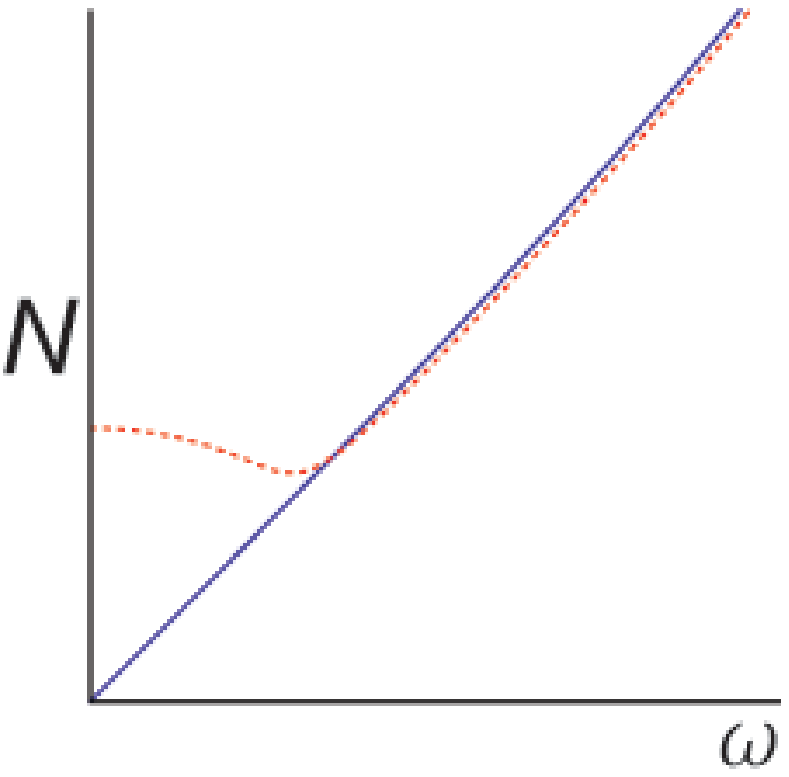} &
\includegraphics[width=29mm]{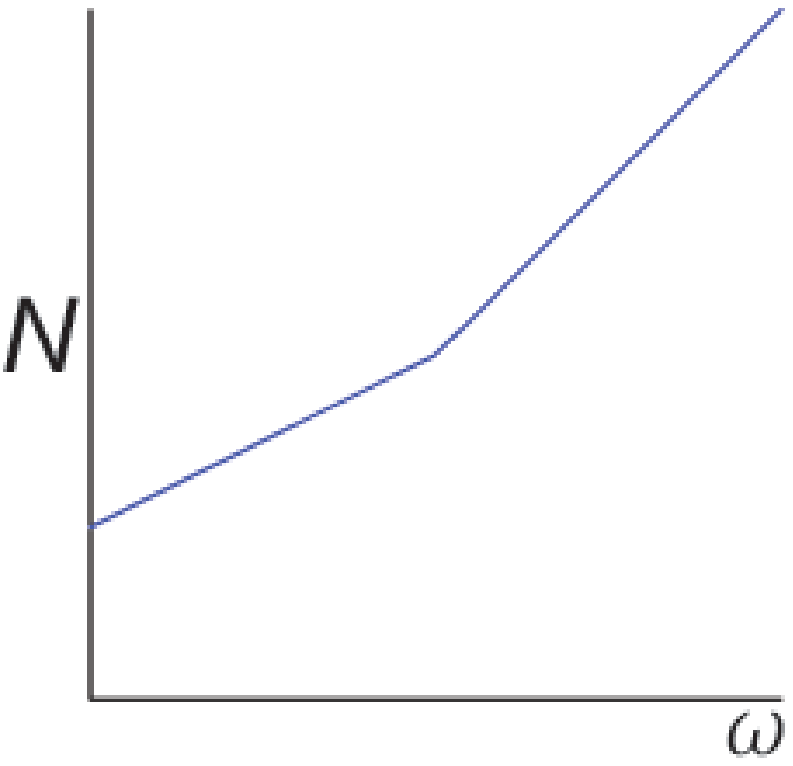} &
\raisebox{-1.5mm}{\includegraphics[width=29mm]{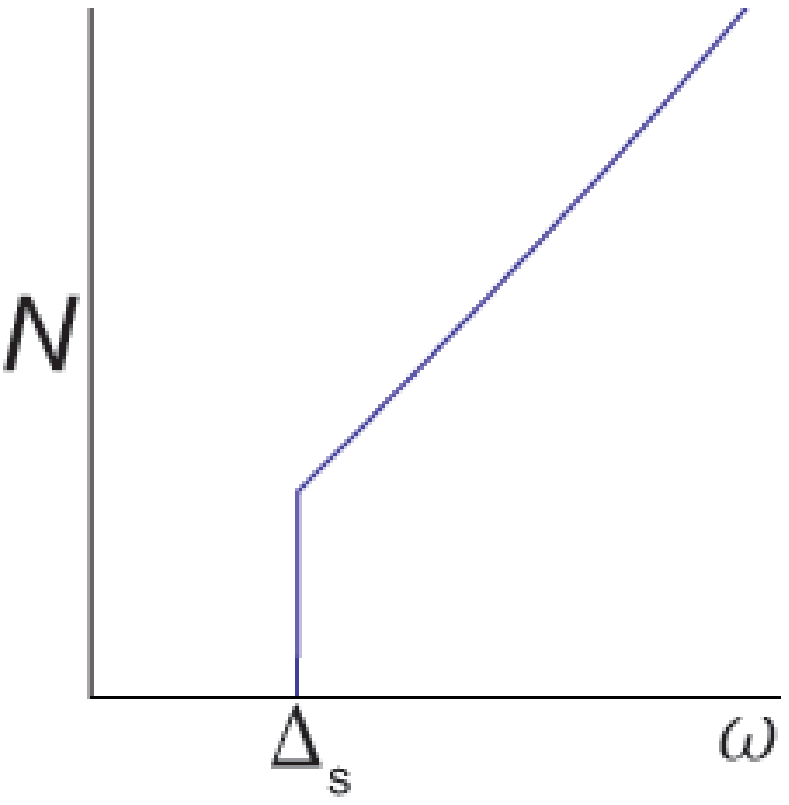}} &
\raisebox{-1.5mm}{\includegraphics[width=29mm]{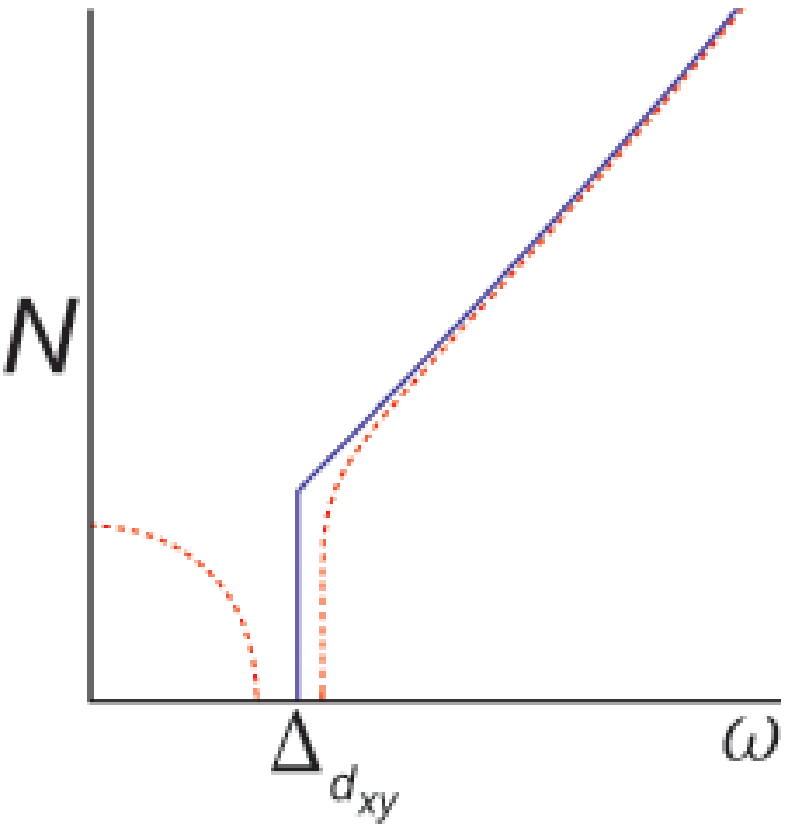}} &
\raisebox{-1.5mm}{\includegraphics[width=29mm]{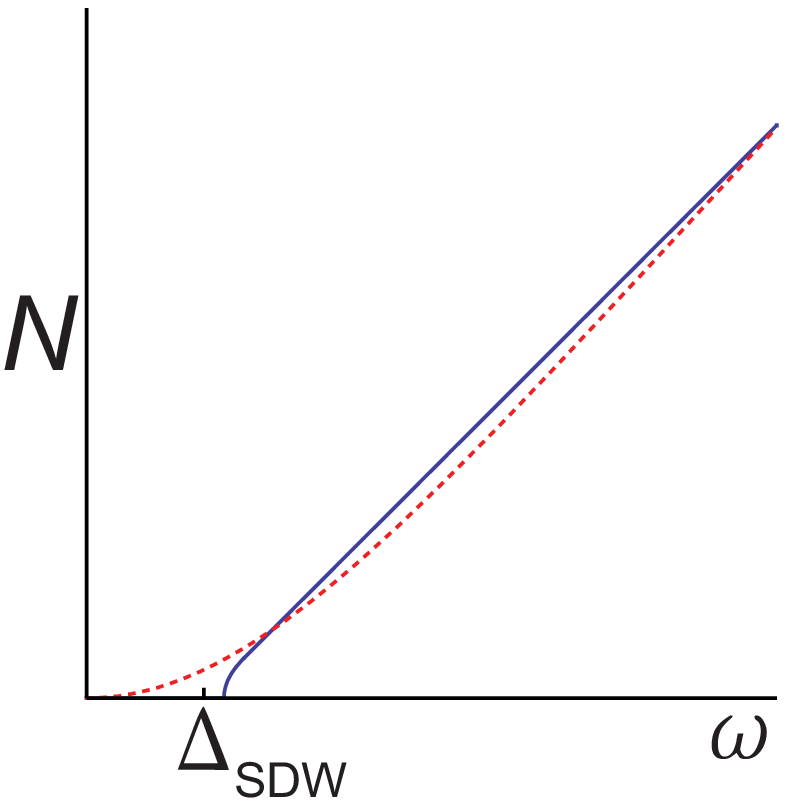}}  \\ 
\hline

\multicolumn{2}{|c|}{\raisebox{17 mm}{\renewcommand{\arraystretch}{1.0}\begin{tabular}[t]{c}
$\rho_s(T)$\\
\textcolor{blue}{clean -- solid}\\
\textcolor{red}{dirty -- dashed}\\
\end{tabular}}}&
 \rule[0mm]{0mm}{32mm} \includegraphics[width=29mm]{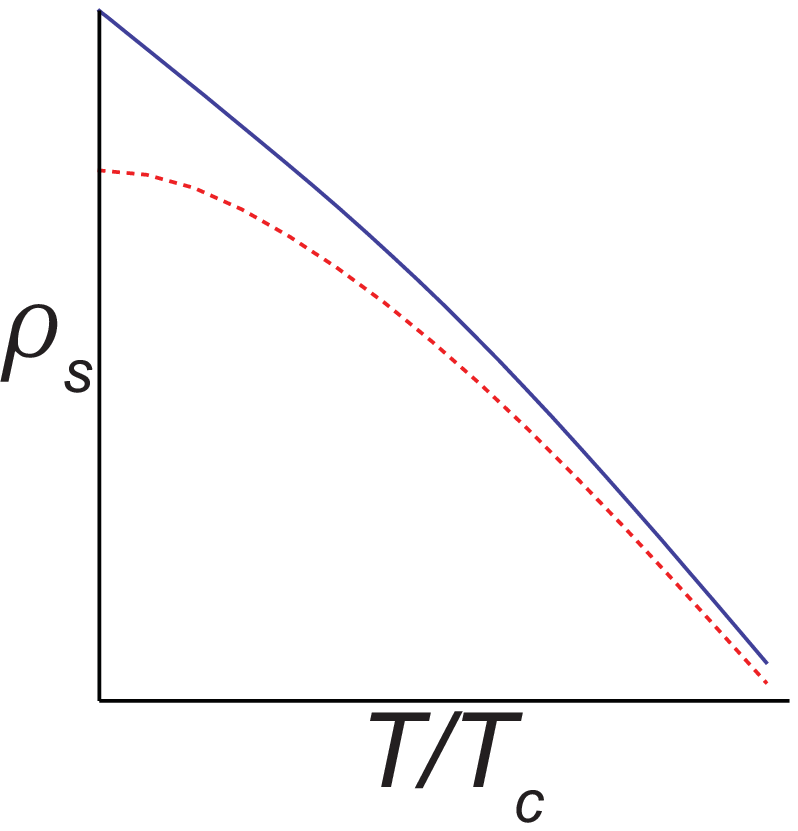} &
\includegraphics[width=29mm]{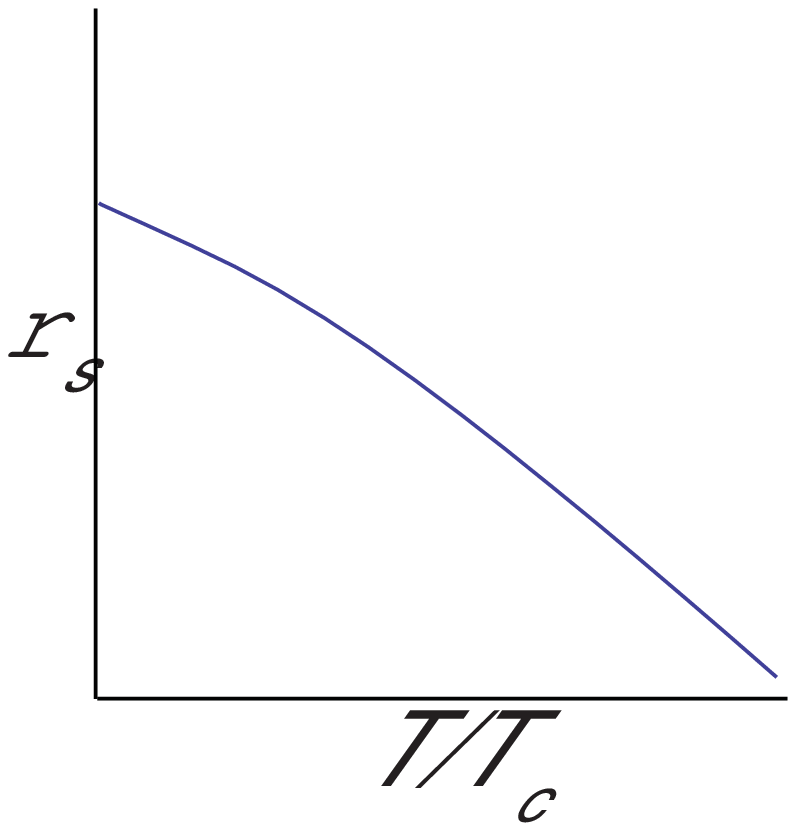} &
\includegraphics[width=29mm]{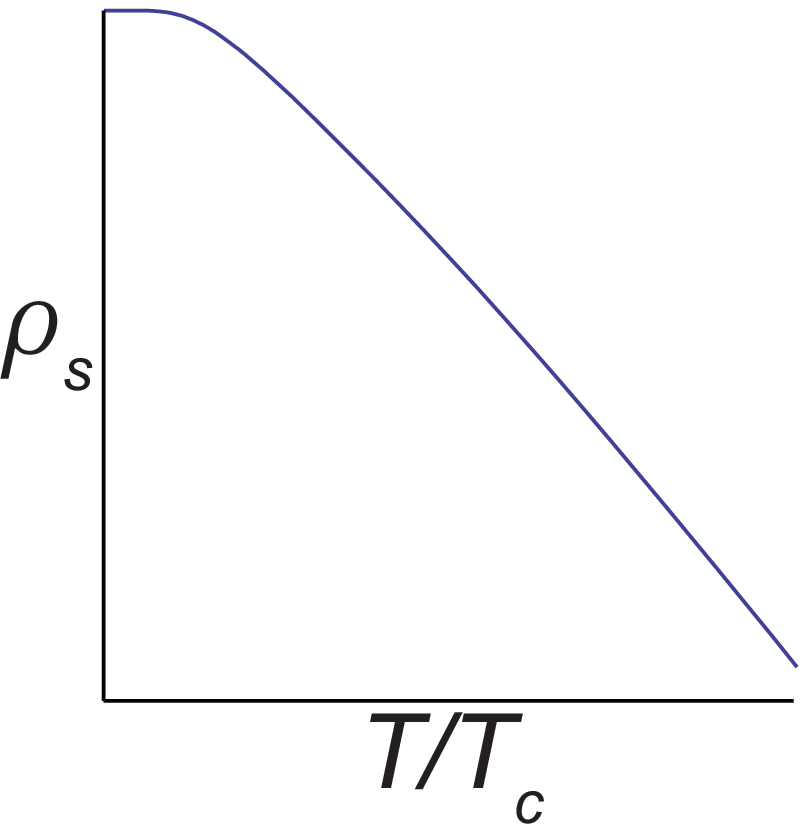} &
\includegraphics[width=29mm]{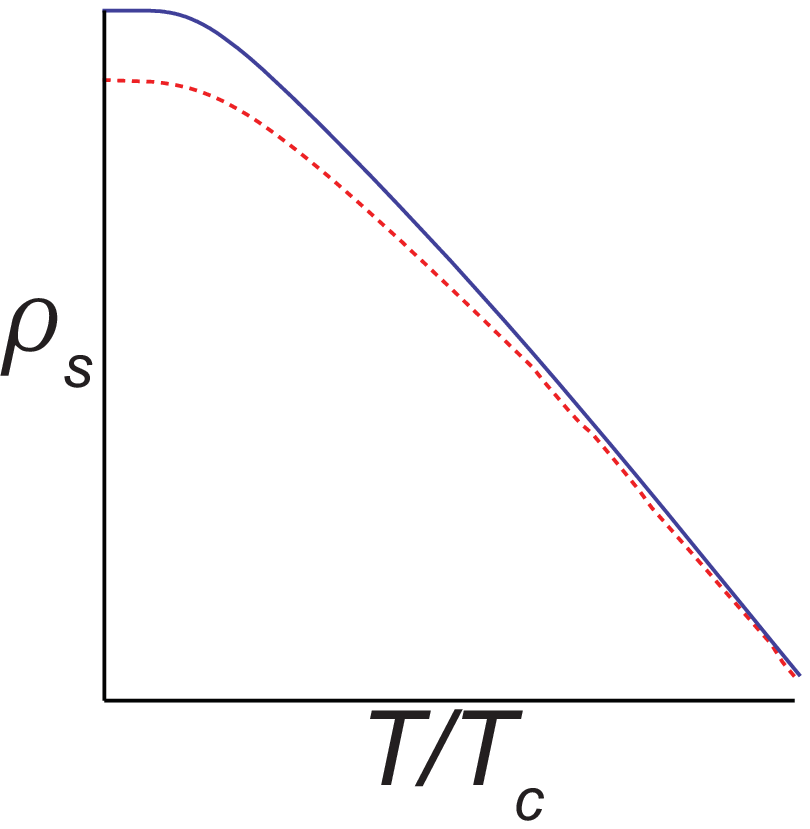} &
\includegraphics[width=29mm]{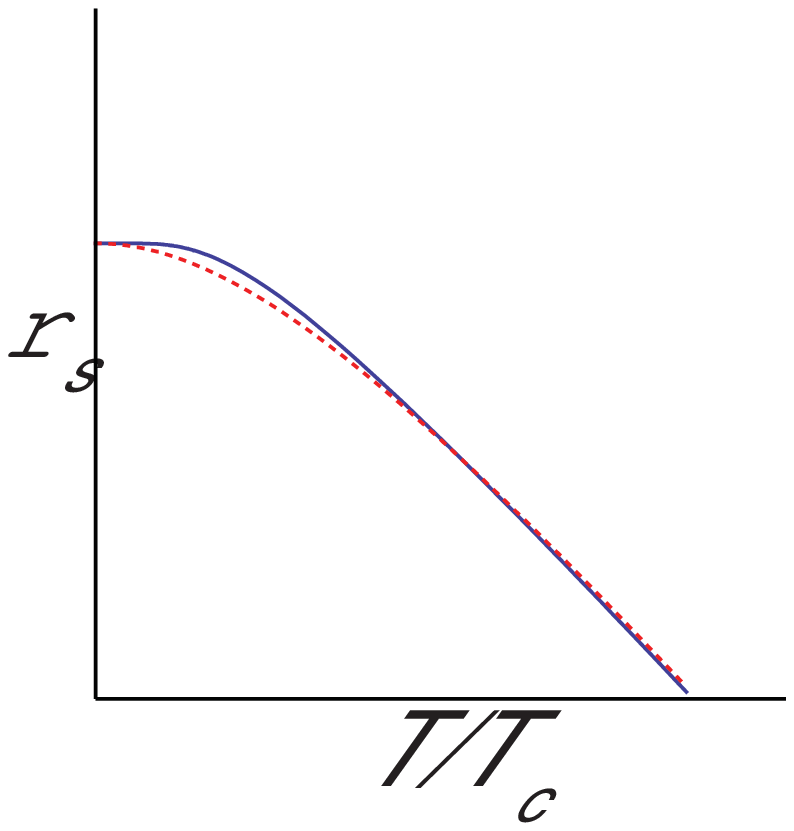} \\  
\hline

\multirow{2}{*}{\renewcommand{\arraystretch}{1.0}\raisebox{-0.5mm}{\begin{tabular}[t]{c} residual \\DOS\\\end{tabular}}}
&  \rule[-1mm]{0mm}{5mm}  clean & No & Yes & No & No & No  \\
& \rule[-2mm]{0mm}{5mm} dirty & Yes & Yes & No & Yes & No  \\ 
\hline

\multirow{2}{*}{\renewcommand{\arraystretch}{1.0}\begin{tabular}[t]{c} low $T$ \\ $\Delta \rho_s(T)$ \\ \end{tabular}}
 & \rule[-1mm]{0mm}{5mm} clean & $T$ & $T/2$ & $\E^{-\Delta_s/k_B T}$ & $\E^{-\Delta_{d_{xy}}/k_B T}$&
$T\E^{-\Delta_\mathrm{SDW}/k_B T}$  \\
 & \rule[-2mm]{0mm}{4mm} dirty & $T^2$ & $T^2$ & $\E^{-\Delta_s/k_B T}$ & $T^{2}$ & $T^{2}$ \\ 
 \hline
 
\end{tabular}
\caption{(color online). Effect of competing orders on the superfluid density of a $d_{x^2 - y^2}$ superconductor.  The first column shows results for the pure $d_{x^2 - y^2}$ state.  Subsequent columns show the effect of competition from: $\Theta_{II}$-type circulating currents;\cite{varma06,berg08} $d_{x^2 - y^2} + \I s$ superconductivity;\cite{modre98} $d_{x^2 - y^2} + \I d_{xy}$ superconductivity; \cite{sharapov06} and spin density waves (SDW) that nest the nodal points.\cite{sharapov06,atkinson07}  The first row shows clean-limit excitation spectra for the near-nodal quasiparticles.  The second row gives the density of states (DOS) $N(\omega)$, both for clean systems and in the presence of disorder.  Note that the effect of disorder has not been calculated for the $\Theta_{II}$-type perturbation, and that nonmagnetic disorder has essentially no effect on the $d_{x^2 - y^2} + \I s$ superconductor.  The third row shows the temperature dependence of the superfluid density $\rho_s(T)$, including deviations from full condensation as $T \to 0$ due to the presence of zero-energy quasiparticles.  The fourth row indicates whether a residual density of states is expected to be seen in $\sigma_1(\Omega,T \to 0)$, and the fifth row gives the leading low temperature behaviour of the superfluid density.  Details of the calculations are given in Appendix~\ref{appendix} for the $d_{x^2 - y^2}$, $d_{x^2 - y^2} + \I s$ and $d_{x^2 - y^2} + \I d_{xy}$ states.  Dirty limit calculations have been made for unitarity limit scatterers.}\label{competingordertable}
\end{table*}

\section{Experiment}\label{experiment}

\begin{figure}[t]
\begin{center}
\includegraphics[width=70mm]{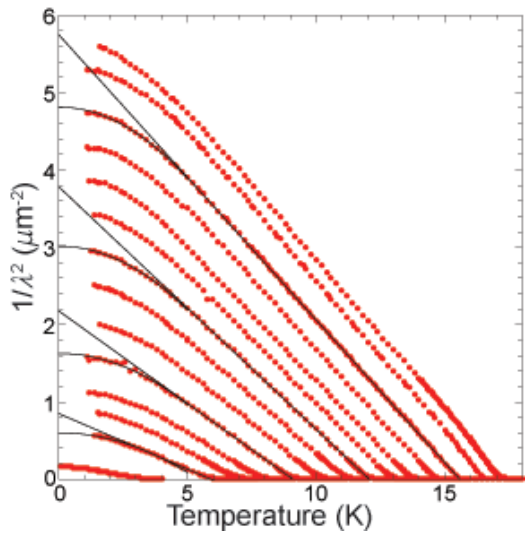}
\caption{(color online). $ab$-plane superfluid density $\rho_s(T)=\lambda^2(T)$ shown at 13 of 37 dopings measured in this study. The straight lines are linear fits the data between 5~K and \tc.  The curved lines are a quadratic fit below 4~K. } 
\label{rhos}
\end{center}
\end{figure}

Measurements of $\rho_s(T)$ and $\sigma_1(\Omega,T)$ have been made on a single-crystal ellipsoid of \ybcou, prepared as described in Ref.~\onlinecite{broun07}.  Following high pressure annealing under a hydrostatic pressure of 35~kbar, controlled relaxation of oxygen order in the CuO chains has been used to continuously tune $T_c$ in the range 17~K to 3~K.  Broadband microwave spectroscopy was carried out early in the sequence, for $T_c = 15.6$~K.  Measurements of $\rho_s(T)$ in the milliKelvin range were made in the fully relaxed state, where $T_c = 3$~K.  

$\rho_s(T)$ is obtained from 2.64~GHz surface impedance measurements, as described in Refs.~\onlinecite{broun07} and \onlinecite{huttema06}.  The sample is positioned at the $H$-field antinode of the TE$_{01\delta}$ mode of a rutile dielectric resonator, with the microwave $H$-field oriented along the $c$ axis of the ellipsoid to induce $ab$-plane screening currents.  Surface impedance $Z_s=R_s+\I X_s$ is obtained using the cavity perturbation approximation: 
\begin{equation}
R_s+\I \Delta X_s = \Gamma\left\{\Delta f_B(T)-2 \I\Delta f_0(T)\right\},
\end{equation}
where $\Delta f_B(T)$ is the change in bandwidth of the TE$_{01\delta}$ mode upon inserting the sample into the cavity; $\Delta f_0(T)$ is the shift in resonant frequency upon warming the sample from base temperature to $T$; and $\Gamma$ is an empirically determined scale factor.  The absolute reactance is set by shifting $\Delta X_s(T)$ so that it matches $R_s(T)$ in the normal state.  We expect local electrodynamics to be a good approximation, giving  $\sigma=\sigma_1-\I\sigma_2=\I\omega\mu_0/Z_s^2$ for the microwave conductivity.  The superfluid density is defined to be $\rho_s\equiv 1/\lambda^2=\omega\mu_0\sigma_2$.

Broadband spectroscopy of the quasiparticle conductivity $\sigma_1(\Omega,T)$ has been carried out using bolometric measurements of $R_s(\Omega,T)$ between 0.1 and 20~GHz, as described in Refs.~\onlinecite{turner03} and \onlinecite{turner04}.  The \ybcou\ ellipsoid and a Ag:Au reference sample were positioned in symmetric locations at the end of a rectangular coaxial transmission line, with the microwave $H$-field again oriented along the $c$ axis of the ellipsoid.  $R_s(\Omega,T)$ has been inferred from the synchronous rise in sample temperature in response to incident microwave fields modulated at 1~Hz.  The Ag:Au sample acts a power meter, providing an absolute calibration.  At low frequencies, $\sigma_1$ can be obtained from $R_s$ from a knowledge of the penetration depth: in this limit $\sigma_1 \approx 2 R_s/\Omega^2 \mu_0^2 \lambda^3$.  At higher frequencies, the quasiparticle conductivity starts to contribute to electromagnetic screening, effectively reducing $\lambda$.  The quasiparticle shielding effect must be taken into account self-consistently, and the procedure for doing this is described in detail in Appendix~C of Ref.~\onlinecite{turner04}.  As part of this process, the quasiparticle contribution to $\sigma_2$ is inferred from a Kramers--Kr\"onig transform of $\sigma_1(\Omega)$.  This in turn requires a robust means of extrapolating $\sigma_1(\Omega)$ outside the measured frequency range.  In previous work,\cite{turner03,ozcan06} we have shown that the phenomenological form,
\begin{equation}
\sigma_1(\Omega) = \sigma_0/[1 + (\Omega/\Gamma)^y]\;,\label{phenomenological}
\end{equation}
 works well for cuprate superconductors, with the exponent $y$ ranging from 1.4 to 1.7. A Drude model, on the other hand, corresponds to $y = 2$.  Physically, the non-Drude exponents stem from the strong energy dependence of scattering rate in an unconventional superconductor.    At low temperatures, thermally excited quasiparticles make a relatively small contribution to electromagnetic screening, so the extraction of $\sigma_1(\Omega)$ from $R_s(\Omega)$ is not particularly sensitive to variations in $y$.  A similar procedure is used to estimate the quasiparticle conductivity spectral weight: in that case there is more sensitivity to the choice of exponent when integrating $\sigma_1(\Omega)$.

\begin{figure}[t]
\begin{center}
\includegraphics[width=70mm]{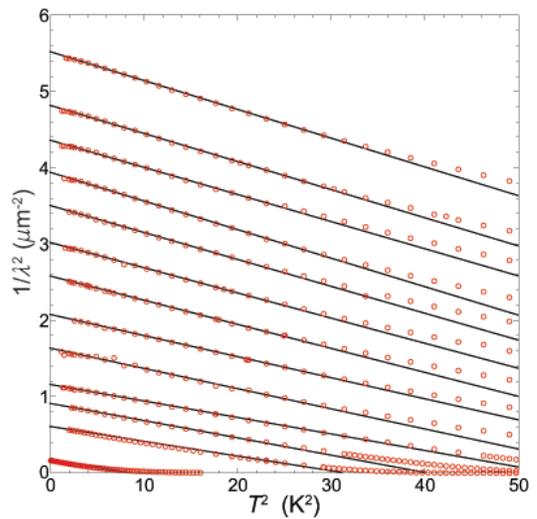}
\caption{(color online). $\rho_s(T)$ plotted versus $T^2$. The straight lines are quadratic fits to the data below 4~K except in the case of the lowest doping ($T_c = 3$~K), where the fit is to just below \tc. The data are linear in $T^2$ up to $T \approx 5$~K.} 
\label{rhosT^2}
\end{center}
\end{figure}

\section{Results and Discussion}\label{results}

\begin{figure}[t]
\begin{center}
\includegraphics[width=55mm]{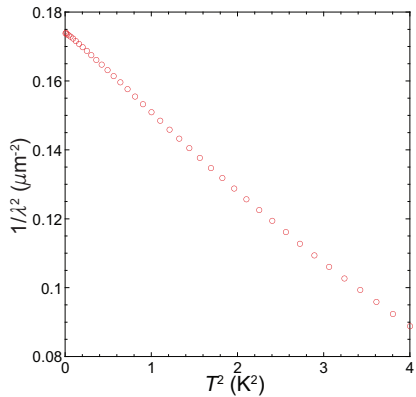}
\caption{(color online). For the lowest doping in this study ($T_c = 3$~K), $\rho_s(T)$ has been measured down to $T = 0.05$~K.  The data, plotted versus $T^2$, reveal that the asymptotic low $T$ behaviour is quadratic in temperature.}
\label{mKrhos}
\end{center}
\end{figure}

$\rho_s(T)$ is plotted in Fig.~\ref{rhos} for a subset of the dopings.  The most prominent feature of the data is the linear $T$ dependence of  $\rho_s$ in the middle of the temperature range, which crosses over to a weaker temperature dependence at low $T$.  The main questions about these data are: what is the limiting low temperature form of  $\rho_s(T)$?; is the crossover the result of disorder?; and is the linear $T$ dependence at higher temperatures characteristic of the behaviour of the ideal, clean system?  To address these issues, we first look at the low temperature range in more detail.  Fig.~\ref{rhosT^2} plots the data from Fig.~\ref{rhos} vs.\ $T^2$, showing that $\rho_s(T)$ indeed crosses over to accurately quadratic behaviour.  For the lowest doping (the fully relaxed state with $T_c \approx 3$~K), the sample has been remounted in our dilution refrigerator system and measured down to $T = 0.05$~K.  This data is plotted vs. $T^2$ in Fig.~\ref{mKrhos}.  We see that the quadratic behaviour is robust to the lowest temperatures, neither flattening out to activated behaviour nor turning up to reveal a power law intermediate between $T^1$ and $T^2$.  

\begin{figure}[t]
\begin{center}
\includegraphics[width=60mm]{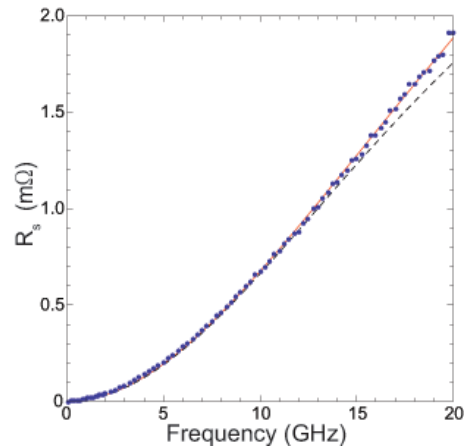}
\caption{(color online). Broadband bolometric measurement of the surface resistance, $R_s(\Omega)$, at $T = 1.7$~K.  Data are for a doping state with $T_c = 15.6$~K.   The solid line is a fit using the phenomenological conductivity model, Eq.~\ref{phenomenological}, with $y = 1.7$.  A fit with $y = 1.4$ is practically indistinguishable and provides an equally good representation of the data.  The dashed line, a best fit to the Drude model ($y = 2$), shows clear deviations at high frequencies.} 
\label{rsdatafit}
\end{center}
\end{figure}

\begin{figure}[h]
\begin{center}
\includegraphics[width=60mm]{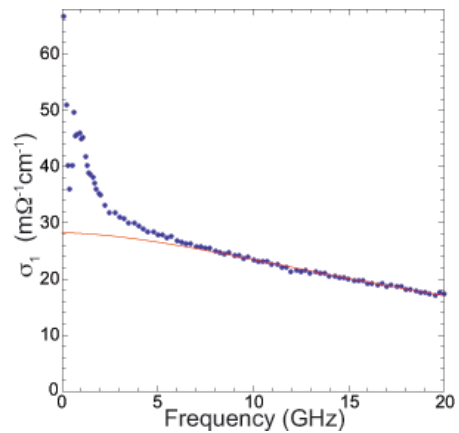}
\caption{(color online). The real part of the conductivity spectrum determined from the $R_s(\Omega)$ data in Fig.~\ref{rsdatafit}.  The solid line is a fit to the conductivity spectrum for $y=1.7$ using the phenomenological model Eq.~\ref{phenomenological}.  The small, narrow peak at low frequencies is a robust result of the analysis and indicates long lived currents, possibly associated with superconducting fluctuations.} 
\label{sigma1fit17}
\end{center}
\end{figure}

To test whether the curvature is the result of disorder, we switch now to broadband microwave spectroscopy, which probes the spectral weight of the zero-energy quasiparticles. Fig.~\ref{rsdatafit} shows $R_s(\Omega)$ at $T = 1.7$~K for the $T_c = 15.6$~K doping.  This has been converted to conductivity $\sigma_1(\Omega)$ in Fig.~\ref{sigma1fit17}, using the self-consistent procedure described in the previous section.  As mentioned above, we use a phenomenological form to fit to the conductivity: $\sigma_1(\Omega) = \sigma_0/[1 + (\Omega/\Gamma)^y]$.  Spectra with $y = 1.4$ and $y=1.7$ provide equally good fits to the $R_s(\Omega)$ data in Fig.~\ref{rsdatafit} --- a Drude fit ($y=2$), however, shows marked deviations at the high frequency end.  At low frequency there is a narrow peak in $\sigma_1(\Omega)$, of uncertain origin, that may be a fluctuation effect.  In any case we are content to omit it from the fitting procedure as it contains an insignificant fraction of the total oscillator strength.  Using the phenomenological model of conductivity, we calculate the uncondensed spectral weight, for different choices of exponent.  Expressed in superfluid density units, we obtain $\Delta \rho_s = 1.05~\mu$m$^{-2}$ for $y = 1.4$ and $\Delta \rho_s = 0.70~\mu$m$^{-2}$ for $y = 1.7$.  Fig.~\ref{rhos} also shows linear and quadratic fits to $\rho_s(T)$ at low temperature.  For comparison with the integrated $T = 1.7$~K spectral weight in $\sigma_1(\Omega)$, we should use the difference between the linear extrapolation of $\rho_s$ to $T = 0$, and $\rho_s(T = 1.7~\mathrm{K})$: this is $\Delta \rho_s = 1.03~\mu$m$^{-2}$.  As this falls within the range estimated from integrating $\sigma_1(\Omega)$, we conclude that the crossover to $T^2$ behaviour in $\rho_s(T)$ is most likely a disorder effect in an otherwise pure $d_{x^2 - y^2}$ state, and that linear fits to $\rho_s(T)$ in the middle of the temperature range should provide a good measure of the low temperature slope in the absence of disorder.

\begin{figure}[t]
\begin{center}
\includegraphics[width=70mm]{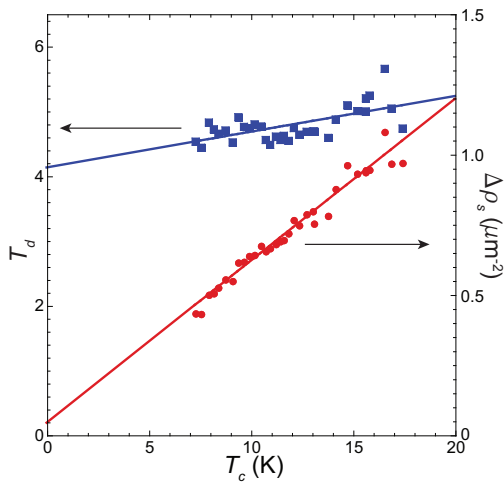}
\caption{(color online). The doping dependence of the disorder crossover temperature $T_d$ and the uncondensed superfluid density $\Delta \rho_s$.  $T_d$ is the temperature at which linear and quadratic fits to $\rho_s(T)$ match in slope, as defined in the text.  $\Delta \rho_s$ is the uncondensed spectral weight predicted from the difference between linear and quadratic extrapolations of $\rho_s(T)$ to $T = 0$, and is consistent with the residual conductivity spectral weight directly measured at $T_c = 15.6$~K via broadband spectroscopy.  } 
\label{Tddeltarhos}
\end{center}
\end{figure}

A useful characterization of the strength of disorder is provided by the temperature \td\ at which $\rho_s(T)$ crosses over from quadratic to linear behaviour. Using an interpolation formula, $\Delta \rho_s(T) = A T^2/(T + 2 T_d)$, similar to that of Ref.~\onlinecite{hirschfeld93a}, \td\ is defined to be the point at which the slope of the  high temperature linear behaviour, $\Delta \rho_s = \alpha T$, matches the slope of the low temperature quadratic behaviour, $\Delta \rho_s = \beta T^2$.  Using values of $\alpha$ and $\beta$ obtained from fits similar to those shown in Figs.~\ref{rhos} and \ref{rhosT^2}, we plot $T_d \equiv \alpha/2 \beta$ in Fig.~\ref{Tddeltarhos}.  The crossover temperature lies between 4~K and 5~K at these low dopings.  This is larger than the cross-over temperature in the best samples of Ortho-II \ybcoOII\ and Ortho-I \ybcoOI, where $T_d$ is less than 1~K.  This is consistent with the lower degree of CuO chain order in \ybcou, which is known to be the dominant source of residual scattering in the best \ybco\ samples,\cite{bobowski06} and is likely enhanced by proximity to the Mott insulator.  Also shown in Fig.~\ref{Tddeltarhos} is the residual DOS, expressed in superfluid density units as $\Delta \rho_s$, and inferred from the difference between linear and quadratic extrapolations of $\rho_s(T)$ to $T = 0$.  $\Delta \rho_s$ falls on underdoping, but remains a roughly constant fraction of $\rho_s(T = 0)$, consistent with the weak doping dependence of $T_d$.

We are able to draw tight conclusions from these measurements about the types and magnitudes of electronic order than might be competing with pure $d_{x^2 - y^2}$ superconductivity in \ybco.  We emphasize that to do this it is essential to have measurements of both the asymptotic low temperature form of $\rho_s(T)$, and the residual DOS from $\sigma_1(\Omega)$.  On the basis of the limiting quadratic $T$ dependence, which we have followed down to $0.05$~K, we can rule out any of the clean-limit behaviours shown in Table~\ref{competingordertable}, as well as the $d_{x^2 - y^2} + \I s$ state in the presence of disorder.  We can also exclude the BCS--BEC crossover scenario, which predicts a $T^{3/2}$ term in $\rho_s(T)$ from incoherent Cooper pairs excited from the condensate.  When disorder is included, four of the remaining states in Table~\ref{competingordertable} are compatible with quadratic behaviour in $\rho_s(T)$.   Of these, nested spin and charge density waves can immediately be eliminated, as they are not expected to be accompanied by a residual DOS.  Of the remaining three, the simplest possibility is pure $d_{x^2 - y^2}$ superconductivity in the presence of a small amount of strong scattering disorder.  However, we cannot rule out a small $\I d_{xy}$ component, nor a weak $\Theta_{II}$-type circulating current phase.  Nevertheless, we can place tight limits on the size of such effects.  We show in Fig.~\ref{disordercrossover} that the $\I d_{xy}$ state only becomes visible once $\Delta_{d_{xy}} > k_B T_d$.  Similarly, we would expect the clean-limit behaviour of the $\Theta_{II}$ state to be apparent once $4 \Delta_\mathrm{cc} > k_B T_d$, meaning that if a perturbation of the form Eq.~\ref{thetaII} is present, then $\Delta_\mathrm{cc}$ must be 1~K or less.\footnote{Although the effect of disorder on $\rho_s$ in the $\Theta_{II}$ state has not been calculated, a rigorous upper bound is set by the residual density of states, which is observed to be about 15\% from broadband quasiparticle spectroscopy of $T_c = 15.6$~K material.  Conservatively assigning all of this to circulating current effects, we would have \mbox{$\Delta \rho_s/\rho_s(T=0) = 0.15 \approx 2 \Delta_\mathrm{cc}/\Delta_0 \approx \Delta_\mathrm{cc}/k_B T_c$}, implying $\Delta_\mathrm{cc} \lesssim 2.4$~K.}  The constraints become even tighter in Ortho-II \ybcoOII\ and Ortho-I \ybcoOI, where the disorder scale $T_d$ is less than 1~K.  

Finally, while we can rule out nested spin and charge density waves, our data say very little about commensurate orders that connect parts of the Fermi surface \emph{away} from the nodes, as these will generally not alter the low energy spectrum.  One such a scenario has been revealed by recent STM measurements on \bscco, \cite{kohsaka08} which show ordered, nondispersing modulations of the DOS at high energies and simultaneously, at low energies, arcs of Bogoliubov quasiparticles associated with the nodal $d_{x^2 - y^2}$ spectrum.  The `Bogoliubov arcs' appear to terminate on the Bragg plane joining $(0,\pi)$ and $(\pi,0)$ points, leaving the nodal spectrum intact.  This would be compatible with the conclusions we draw here about \ybco.

\section{Conclusions}\label{conclusions}

We have shown that measurements of superfluid density can be used as a sensitive probe of electronic orders than might compete with pure $d_{x^2 - y^2}$ superconductivity.  Broadband conductivity measurements provide complementary information on zero-energy quasiparticles that would be difficult to infer from $\rho_s(T)$ alone.  Measurements on underdoped \ybcou\ reveal a crossover from linear to quadratic behaviour in $\rho_s(T)$ below a temperature $T_d \approx $~4~K to 5~K.  The $T^2$ power law has been followed as low as 0.05~K and appears to be the asymptotic low temperature behaviour.  It is also accompanied by a residual quasiparticle spectral weight of corresponding magnitude, leading us to conclude that the crossover is a disorder effect.  The observations immediately allow us to rule out BCS--BEC crossover physics; competition from $d_{x^2 - y^2} + \I s$ superconductivity; and spin and charge density waves that nest the nodal points.  Due to the presence of disorder, we cannot eliminate the possibility of either disordered $d_{x^2 - y^2} + \I d_{xy}$ superconductivity, provided $\Delta_{d_{xy}} \lesssim 4$ -- 5~K; or a perturbation of the form Eq.~\ref{thetaII} from a $\Theta_{II}$-type circulating current phase, as long as $\Delta_\mathrm{cc} \lesssim 1$~K.  The small magnitude of  the term is compatible with related observations from $\mu$SR,\cite{sonier01} neutron scattering\cite{fauque06} and polar Kerr-effect measurements.\cite{xia08}

\acknowledgements

We would like to thank J.~Carbotte, P.~J.~Hirschfeld, S.~Kivelson and J.~E.~Sonier for useful discussions. This work was funded by the National Science and Engineering Research Council of Canada and the Canadian Institute for Advanced Research.\\

\appendix

\section{\dwave, $d_{x^2 - y^2} + \I d_{xy}$ and  $d_{x^2 - y^2} + \I s$ states}\label{appendix}

The $d_{x^2 - y^2} + \I d_{xy}$ state and the  $d_{x^2 - y^2} + \I s$ state are two candidate order parameters that may compete with pure  $d_{x^2 - y^2}$ superconductivity in the cuprates.  In this appendix we review the theory of the penetration depth in the presence of disorder and gauge the extent to which these states can be distinguished by microwave experiments.  The theory of unconventional superconductivity in the presence of elastic scattering disorder has been developed by many authors,\cite{nam67,pethick86,hirschfeld88,prohammer91,schachinger03,hirschfeld93,hirschfeld93a,borkowski94,hirschfeld94} and has been reviewed in several places.\cite{joynt97,hussey02,balatsky06} In these systems, disorder not only imparts a finite lifetime to the quasiparticles, it alters the excitation spectrum by pair-breaking, and the two effects must be dealt with together.  The self-consistent $t$-matrix approximation (SCTMA) provides a powerful approach for capturing this physics, particularly in the resonant scattering limit, where the impurity is on the verge of binding a quasiparticle at the Fermi energy.  In the SCTMA, impurities are usually approximated as point defects that scatter in the $s$-wave channel.  The effect of the disorder is to renormalize the quasiparticle energy $\omega$ and the superconducting gap $\Delta_\mathbf{k}$, which can be expressed in the following way:
\begin{align}
\omega \to \tilde\omega & = \omega + \I \pi \Gamma \frac{N(\omega)}{c^2 + N^2(\omega) + P^2(\omega)}\label{renorm1}\\
\Delta_\mathbf{k} \to \tilde\Delta_\mathbf{k} & = \Delta_\mathbf{k} + \I \pi \Gamma \frac{P(\omega)}{c^2 + N^2(\omega) + P^2(\omega)}\label{renorm2}\;.
\end{align}
Here $\Gamma = n_i n/\pi^2 D(\epsilon_F)$, where $n_i$ is the impurity concentration, $n$ is the conduction electron density, and $D(\epsilon_F)$ is the density of states at the Fermi level.\cite{hirschfeld93a}  The impurity scattering strength is characterized by $c$, the cotangent of the $s$-wave scattering phase shift.   The quasiparticle density $N(\omega)$ and pair density $P(\omega)$ depend on details of the particular superconducting state and are defined below for the different types of order parameter.  
For purely unconventional order parameters, $\langle \Delta_\mathbf{k} \rangle_\mathrm{FS} = 0$ and $P(\omega)$ vanishes ---  these states are therefore  unrenormalized by $s$-wave scatterers.

We are primarily interested in the behaviour of the low energy excitations so, without loss of generality, we take the two-dimensional Fermi surface to be isotropic, and the gap functions to be the simplest cylindrical harmonics of the required symmetry:
\begin{align}
\Delta_{d_{x^2 - y^2}} & = \Delta_0 \cos 2 \phi\;,\\
\Delta_{d_{xy}} & = \eta \Delta_0 \sin 2 \phi\;,\\
\Delta_s & = \zeta \Delta_0\;.
\end{align}
Here $\phi$ measures angle from the Cu--O bond direction and $\eta$ and $\zeta$ are constants.  For the pure $d_{x^2 - y^2}$ state there is no gap renormalization.  The quasiparticle density is
\begin{equation}
N(\omega)  = \left\langle \frac{\tilde\omega}{\sqrt{\tilde\omega^2 - \Delta_0^2 \cos^2 2 \phi}}\right\rangle_\phi = \frac{2}{\pi} K\!\left(\frac{\Delta_0^2}{\tilde\omega^2} \right)\;,\label{dos}
\end{equation}
where $\langle ... \rangle_\phi$ is an angle average around the cylindrical Fermi surface, $K(x)$ is the complete elliptic integral of the first kind, and the branch of the square root in Eq.~\ref{renorm1} is chosen so that $\tilde \omega$ has positive imaginary part.  In the strong-scattering (unitarity) limit, for instance, $c = 0$ and $\tilde\omega(\omega)$ is a root of
\begin{equation}
\tilde\omega = \omega + \I \pi^2 \Gamma/2 K(\Delta_0^2/\tilde\omega^2).
\end{equation}
$\tilde{\omega}(\omega)$ encodes all the physics of scattering and pair-breaking. Inserted into the real part of Eq.~\ref{dos} it gives the quasiparticle density of states in the presence of disorder.  To calculate penetration depth using $\tilde \omega$, a modification of Eq.~\ref{lambdatwo} is used:\cite{hirschfeld93a}
\begin{equation}
\frac{\lambda^2_0}{\lambda^2(T)} =  \tfrac{1}{2} \int_{-\infty}^\infty \!\!\!\!\!\D \omega\, \tanh\frac{\omega}{2 k_B T}\re \left\langle\!\frac{\tilde\Delta_\mathbf{k}^2 }{(\tilde\omega^2 - \tilde\Delta_\mathbf{k}^2)^\frac{3}{2}}\!\right\rangle_{\!\mathrm{FS}}\;.\label{dirtylambda}
\end{equation}
The density of states factor is  
\begin{equation}
\begin{split}
 \left\langle \frac{\tilde\Delta_\mathbf{k}^2 }{(\tilde\omega^2 - \tilde\Delta_\mathbf{k}^2)^\frac{3}{2}}\right\rangle_\mathrm{FS} = \left\langle \frac{\Delta_0^2 \cos^2 2 \phi}{(\tilde\omega^2 - \Delta_0^2 \cos^2 2 \phi)^\frac{3}{2}}\right\rangle_\phi\\
 = \frac{2}{\pi \tilde\omega}\left(K(\Delta_0^2/\tilde\omega^2) + \frac{\tilde\omega^2 }{\Delta_0^2 - \tilde\omega^2}E(\Delta_0^2/\tilde\omega^2)\right),
\end{split}
\end{equation}
where $E(x)$ is the complete elliptic integral of the second kind.

\begin{figure}[t]
\begin{center}
\includegraphics[width=83mm]{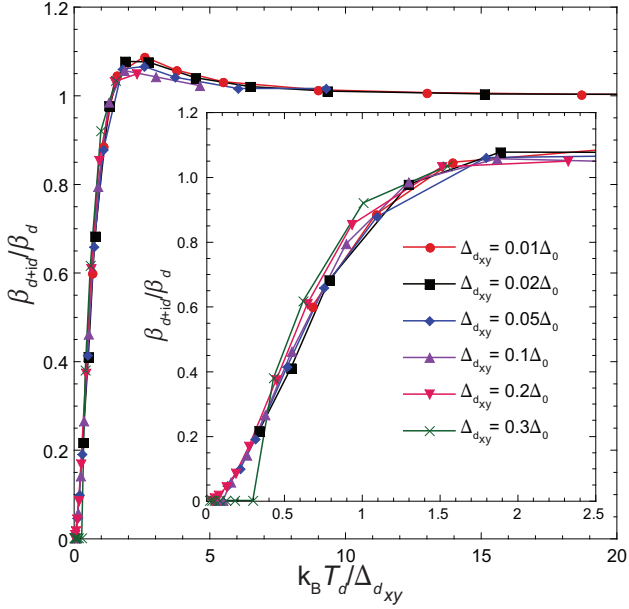}
\caption{(color online).  The onset of quadratic temperature dependence of $\rho_s(T)$ in a $d_{x^2 - y^2} + \I d_{xy}$ superconductor as a function of disorder strength, relative to that of a $d_{x^2 - y^2}$ state.  $\beta_{d + \I d}$ is the $T^2$ coefficient of $\rho_s(T)$ for the $d_{x^2 - y^2} + \I d_{xy}$ superconductor.  $\beta_d$ is the same quantity for the $d_{x^2 - y^2}$ state.  Disorder strength is characterized by the disorder crossover temperature $T_d$ of the $d_{x^2 - y^2}$ superconductor, as defined in the text.  Data are plotted for different values of the $d_{xy}$ gap, $\Delta_{d_{xy}}$, and scale well as a function of $T_d/\Delta_{d_{xy}}$.  The two pairing states are difficult to distinguish on the basis of $\Delta \rho_s(T)$ once $k_B T_d \gtrsim \Delta_{d_{xy}}$. } 
\label{disordercrossover}
\end{center}
\end{figure}

For the $d_{x^2 - y^2} + \I d_{xy}$ state, $\Delta(\phi) = \Delta_0(\cos 2 \phi + \I \eta \sin 2 \phi)$, and there is similarly no gap renormalization.  The quasiparticle density is 
\begin{equation}
\begin{split}
N(\omega) & = \left\langle \frac{\tilde\omega}{\sqrt{\tilde\omega^2 - \Delta_0^2(\cos^2 2 \phi + \eta^2 \sin^2 2 \phi)}}\right\rangle_\phi\\
& = \frac{2}{\pi} \frac{\tilde\omega}{\sqrt{\tilde\omega^2 - \eta^2 \Delta_0^2}}K\!\left(\frac{(1 - \eta^2)\Delta_0^2}{\tilde\omega^2 - \eta^2 \Delta_0^2} \right)\;.
\end{split}
\end{equation}
The density of states factor in Eq.~\ref{dirtylambda} becomes
\begin{equation}
\begin{split}
& \left\langle \frac{\tilde\Delta_\mathbf{k}^2 }{(\tilde\omega^2 - \tilde\Delta_\mathbf{k}^2)^\frac{3}{2}}\right\rangle_\mathrm{FS} \\ 
&= \left\langle \frac{\Delta_0^2 (\cos^2 2 \phi + \eta^2 \sin^2 2 \phi)}{\left(\tilde\omega^2 - \Delta_0^2 (\cos^2 2 \phi + \eta^2 \sin^2 2 \phi)\right)^\frac{3}{2}}\right\rangle_\phi\\
& = \frac{2}{\pi} \frac{1}{\sqrt{\tilde\omega^2 - \eta^2 \Delta_0^2}} \; \times \\ 
& \left[K\!\left(\frac{(1 - \eta^2)\Delta_0^2}{\tilde\omega^2 - \eta^2 \Delta_0^2} \right) \!+ \!\frac{\tilde\omega^2 }{\Delta_0^2 - \tilde\omega^2}E\!\left(\frac{(1 - \eta^2)\Delta_0^2}{\tilde\omega^2 - \eta^2 \Delta_0^2} \right)\right]
\end{split}
\end{equation}
In the $d_{x^2 - y^2} + \I s$ state, impurity renormalization of $\Delta_s$ must be taken into account.  The renormalization equations \ref{renorm1} and \ref{renorm2} can be rewritten
\begin{align}
1 & = \frac{\omega}{\tilde\omega} + \I \pi \Gamma \frac{N(\omega)/\tilde\omega}{c^2 + N^2(\omega) + P^2(\omega)}\label{renorms1}\\
1 & = \frac{\Delta_s}{\tilde\Delta_s} + \I \pi \Gamma \frac{P(\omega)/\tilde\Delta_s}{c^2 + N^2(\omega) + P^2(\omega)}\label{renorms2}\;,
\end{align}
where 
\begin{align}
N(\omega)  & = \left\langle \frac{\tilde\omega}{\sqrt{\tilde\omega^2 - \Delta_0^2 \cos^2 2 \phi - \tilde\Delta_s^2}}\right\rangle_\phi\\
P(\omega)  & = \left\langle \frac{\tilde\Delta_s}{\sqrt{\tilde\omega^2 - \Delta_0^2 \cos^2 2 \phi - \tilde\Delta_s^2}}\right\rangle_\phi\;.
\end{align}
Since $N(\omega)/\tilde\omega = P(\omega)/\tilde\Delta_s$,  the quantities $\omega/\tilde\omega$ and $\Delta_s/\tilde\Delta_s$ obey identical equations and therefore $\tilde\Delta_s =  \Delta_s\tilde\omega/\omega$.  \ref{renorms1} and \ref{renorms2} can then be combined into a single equation
\begin{equation}
\tilde\omega = \omega + \I \pi \Gamma \frac{N(\omega)}{c^2 + N^2(\omega)(1 + \Delta_s^2/\omega^2)}\;,\label{dis1}
\end{equation}
where
\begin{equation}
\begin{split}
N(\omega)  & = \left\langle \frac{\tilde\omega}{\sqrt{\tilde\omega^2(1 - \Delta_s^2/\omega^2) - \Delta_0^2 \cos^2 2 \phi}}\right\rangle_\phi\\
&  = \frac{2}{\pi} \frac{1}{\sqrt{1 - \Delta_s^2/\omega^2}}K\!\left(\frac{\Delta_0^2}{(1 - \Delta_s^2/\omega^2)\tilde\omega^2} \right)\;.\label{dis2}
\end{split}
\end{equation}
The corresponding term in Eq.~\ref{dirtylambda} is
\begin{equation}
\begin{split}
& \left\langle \frac{\tilde\Delta_\mathbf{k}^2 }{(\tilde\omega^2 - \tilde\Delta_\mathbf{k}^2)^\frac{3}{2}}\right\rangle_\mathrm{FS} \\ 
&= \left\langle \frac{\Delta_0^2 (\cos^2 2 \phi + \eta^2 \sin^2 2 \phi)}{\left(\tilde\omega^2 - \Delta_0^2 (\cos^2 2 \phi + \eta^2 \sin^2 2 \phi)\right)^\frac{3}{2}}\right\rangle_\phi\\
& = \frac{2}{\pi} \frac{1}{\tilde \omega\sqrt{1 - \frac{\Delta_s^2}{\omega^2}}} \; \times \\ & \left[K\!\left(\!\frac{\Delta_0^2}{\big( 1\!-\! \frac{\Delta_s^2}{\omega^2}\big)\tilde\omega^2}\! \right) \!+ \!\frac{\tilde\omega^2}{\Delta_0^2 \!\!-\!\! \big( 1\!\!-\!\! \frac{\Delta_s^2}{\omega^2} \big)\tilde\omega^2}E\!\left(\!\frac{\Delta_0^2}{\big( 1\!- \!\frac{\Delta_s^2}{\omega^2} \big)\tilde\omega^2}\! \right)\right]\;.
\end{split}
\end{equation}

We are now in a position to compare results for the three order parameters.  The forms for the density of states $N(\omega)$ and the superfluid density $\rho_s(T)$ are shown in Table~\ref{competingordertable}, both in the clean limit and in the presence of strong scattering disorder ($c$ = 0).  
The key feature of the clean $d_{x^2 - y^2} + \I d_{xy}$ and  $d_{x^2 - y^2} + \I s$ states is a finite energy gap, giving rise to activated behaviour in $\rho_s(T)$.  The $d_{x^2 - y^2} + \I d_{xy}$ and  $d_{x^2 - y^2} + \I s$ states behave very differently in response to disorder.  In the $d_{x^2 - y^2} + \I s$ case, the energy gap is robust.  This can be traced back to the expressions for the renormalized frequency, Eqs.~\ref{dis1} and \ref{dis2}.  Impurity renormalization of $\Delta_s$ leads to solutions for $\tilde \omega$ that are purely real for $\omega < \Delta_s$, preventing the formation of any low-lying quasiparticle states in $N(\omega)$.\cite{borkowski94}  The  $d_{x^2 - y^2} + \I d_{xy}$ case is quite different: pair breaking occurs for even small amounts of disorder, leading immediately to a $T^2$ term in $\rho_s(T)$.  The $T^2$ term starts out weak, but grows in magnitude until it is comparable to that of a pure $d_{x^2 - y^2}$ superconductor with a similar amount of disorder.  This cross over is charted in Fig.~\ref{disordercrossover}, which shows that the $d_{x^2 - y^2}$ and $d_{x^2 - y^2} + \I d_{xy}$ states become indistinguishable when the energy scale for the disorder, $k_B T_d$, becomes comparable to $\Delta_{d_{xy}}$.

\clearpage

%\bibliography{uYBCOdisorder}% Produces the bibliography via BibTeX.

\end{document}